\documentclass[fleqn,usenatbib]{mnras}

\usepackage[T1]{fontenc}

\DeclareRobustCommand{\VAN}[3]{#2}
\let\VANthebibliography\thebibliography
\def\thebibliography{\DeclareRobustCommand{\VAN}[3]{##3}\VANthebibliography}

\newcommand{\dd}{{\rm d}}

\usepackage{hyperref}
\usepackage{graphicx}	
\usepackage{amsmath}	
\usepackage{amssymb}	
\usepackage{tabularx}

\usepackage{subfig}
\usepackage{tabularx}

\title{Tracing Milky Way substructure with an RR Lyrae hierarchical clustering forest}

\author[B.~T. Cook et al.]{Brian~T.~Cook$^{1,2}$\thanks{E-mail: briantcook3070@gmail.com. Distribution A. Approved for public release - distribution is unlimited},
Deborah~F.~Woods,$^{2}$,
Jessica~D.~Ruprecht$^{2}$,
Jacob~Varey$^{2}$,
Radha~Mastandrea$^{3,2}$,
\newauthor
Kaylee~de~Soto$^{3,2}$,
Jacob~F.~Harburg$^{2}$,
Umaa~Rebbapragada$^{4}$,
Ashish~A.~Mahabal$^{5,6}$
\\
$^{1}$Center for Relativistic Astrophysics, School of Physics, Georgia Institute of Technology, Atlanta, GA, 30332, USA \\
$^{2}$MIT Lincoln Laboratory, Lexington, MA, 02421, USA \\
$^{3}$Department of Physics, Massachusetts Institute of Technology, Cambridge, MA, 02142, USA \\
$^{4}$Jet Propulsion Laboratory, California Institute of Technology, Pasadena, CA, 91109, USA \\
$^{5}$Division of Physics, Mathematics and Astronomy, California Institute of Technology, Pasadena, CA, 91125, USA \\
$^{6}$Center for Data Driven Discovery, California Institute of Technology Pasadena, CA, 91125, USA
}

\date{Accepted 2022 April 6. Received 2022 April 6; in original form 2021 22 July}

\pubyear{2022}

\begin{document}
\label{firstpage}
\pagerange{\pageref{firstpage}--\pageref{lastpage}}
\maketitle

\begin{abstract}
RR Lyrae variable stars have long been reliable standard candles used
to discern structure in the Local Group. With this in mind, we present a routine to identify groupings containing a statistically significant number of RR Lyrae variables in the Milky Way environment. RR Lyrae variable groupings, or substructures, with potential Galactic archaeology applications are found using a forest of agglomerative, hierarchical clustering trees, whose leaves are Milky Way RR Lyrae variables. Each grouping is validated by ensuring that the internal RR Lyrae variable proper motions are sufficiently correlated. Photometric information was collected from the {\it Gaia} second data release and proper motions from the (early) third data release. After applying this routine to the catalogue of 91234 variables, we are able to report sixteen unique RR Lyrae substructures with physical sizes of less than 1 kpc. Five of these substructures are in close proximity to Milky Way globular clusters with previously known tidal tails and/or a potential connection to Galactic merger events. One candidate substructure is in the neighbourhood of the Large Magellanic Cloud but is more distant (and older) than known satellites of the dwarf galaxy. Our study ends with a discussion of ways in which future surveys could be applied to the discovery of Milky Way stellar streams.
\end{abstract}




\section{Introduction} 
\label{sec:intro}

The Milky Way's stellar population is laced with substructures from which dwarf galaxy merger events and star formation histories can be inferred across cosmic time. The hierarchical galaxy formation picture suggests that galaxies like our own grow with time as orbiting dwarves
are accreted \citep[e.g.,][]{press1974,bullock2005}. Tidal shocking and subsequent stripping occurs during these mergers \citep[see][for a review]{binney2008}; however, disrupted clusters have been successfully identified using stellar configurations in the Milky Way's phase space \citep[e.g.,][]{portegieszwart2009a, malhan2019}. 

Many Galactic archaeological artifacts can be analyzed using RR Lyrae variable stars, a standard candle popular in Local Group studies \citep[see][for a review]{preston1964}. Globular Clusters (GCs), structures of Population II stars that formed very early in a galaxy’s lifetime \citep{searle1978}, are especially critical. The age and metallicity distributions of GCs, for example, can be used to trace the star formation of their host galaxy \citep{kisslerpatig1999}. Globular clusters can also be helpful in determining the Milky Way's merger history \citep[e.g.][]{kruijssen2020, bonaca2021}. RR Lyrae variables are most frequently associated with GCs \citep[e.g.,][]{zinn1985, clement2001}, but their utility has been demonstrated in a variety of Milky Way satellite studies as well. The so-called Pisces
Overdensity, a metal-poor satellite related to the infall of the Large Magellanic Cloud, has been successfully analyzed using an RR Lyrae
variable catalogue \citep[e.g.,][]{watkins2009, belokurov2019}.

RR Lyrae variables have also been found in stellar streams \citep[e.g.,][]{duffau2006, vivas2016}, showing that this variable star class populates Galactic substructure in the form of tidal debris. This property is useful even after the progenitor object has undergone serious tidal distortion. The kinematic space morphology of debris from infalling dwarves can, in turn, be used to make inferences about the Galactic merger history. There are three general tidal debris morphology classes, with early merger events manifesting primarily as ``cloudy" and ``great circle" debris \citep{johnston2008}, the former comprised mostly of stars on highly eccentric orbits. A recent study of RR Lyrae variables classified using {\it Gaia} DR2 \citep{iorio2019} does not use this classification system, but confirms that RR Lyrae variables in the inner Galactic halo can be classified as ``cloudy" debris that were primarily sourced by a single ancient merger event. 

Thus, we are motivated to find substructures in the Milky Way environment comprised of RR Lyrae variable stars, and there is a suite of clustering methodologies to be considered. The easiest set of objects to cluster would have obvious categorical features; if twenty people were distributed between New York, Tokyo, and Buenos Aires, for example, then there would be three clusters found within a data structure tabulating personal separation distances, and we would probably not need to write a computer program to determine the clusters. The distribution of Milky Way RR Lyrae variables, however, is remarkably more complex, so we must rely upon
an automated clustering algorithm to glean RR Lyrae substructures. One of our primary goals is to determine an optimal clustering method such that rich structure can be found
without overfitting to the data. One potential method is a Gaussian Mixture Model that incorporates
an overfitting penalty \cite[such as the AIC, see][]{akaike1998}, but this would be built upon the assumption that the
variable stars are organized into subpopulations that are independently normally distributed. Choosing a number of
clusters or nearest neighbours {\it a priori} and then iteratively optimizing a cost function is another option, but this would
provide no information as to what extent neighbouring clusters are related.

An agglomerative, hierarchical clustering algorithm, when applied to a catalogue of RR Lyrae variables, mitigates
both of these concerns. Agglomerative, hierarchical clustering is the process by which $n$ clusters each containing one object (in
our case, an RR Lyrae variable ``leaf") accumulate in a branch-like fashion into a single ``root" cluster populated by $n$ objects \citep{gower1969, gordon1987, everitt2011}. Hierarchical clustering has been applied in other astronomical contexts, including galaxy
distributions \citep{peebles1974} and analyses of the SDSS “Great Wall” \citep{ivezic2014}. Analyses of hierarchical structures of this kind are demonstrably a powerful method of interpreting a variety of complex networks \citep{clauset2008}. It is worth noting, however, that using the term ``clusters" here is an unfortunate coincidence, as we do not want to unintentionally conflate our findings with open and globular clusters. Confirming a single merger via analyses of the tree's root (i.e., RR Lyrae distribution on the largest scales) is also beyond the scope of this study, as we will discuss in \S \ref{sec:methods}. It is encouraging, nonetheless, that hierarchical clustering trees might help connect the Milky Way's RR Lyrae population to Galactic merger events in a variety of ways.

This study's contributions to the Galactic archaeology literature is two-fold: we present a new way of identifying substructures in the Milky Way using RR Lyrae hierarchical clustering trees (described in \S \ref{subsec:hierarchical_clustering}) whose coverage includes the entire sky plane, as well as substructure candidates suitable for future studies, including a previously unknown mid-halo substructure and potential Large Magellanic Cloud (LMC) satellite. The standardization of photometric and proper motion information, as well as our clustering algorithm, is described and tested in \S \ref{sec:methods}. Each RR Lyrae variable is placed in a five-dimensional phase space (distance modulus + 2D sky plane coordinates + 2D sky plane proper motions) used for clustering and validation. We explore a small portion of the stellar stream parameter space to test our algorithm's effectiveness, and confirm its utility by identifying a known stream of RR Lyrae variables. Our method of identifying RR Lyrae variable substructures and the subsequent results are presented in \S \ref{sec:results}. A discussion in \S \ref{sec:discussion} focuses on the relationship between forest groupings and the Milky Way globular cluster population, as well as potential future studies in which kinematic and spectroscopic information could be leveraged to identify Galactic stellar streams.

\section{Methods}
\label{sec:methods}

\subsection{RR Lyrae Variables as standard candles}
\label{subsec:rr_lyrae_candles}

A well-worn rung on the cosmic distance ladder, as mentioned in \S \ref{sec:intro}, is the RR Lyrae variable star. The variability of the RR Lyrae is due to changes in the stellar opacity, which causes a periodic ebb and flow of the star’s luminosity \citep{maeder2009}. The period of the RR Lyrae variable directly relates to its absolute magnitude, allowing astronomers to calculate absolute distances to stars of this type. Once the absolute magnitude has been computed from the variable’s pulsation period, the luminosity can be inferred using the solar luminosity and bolometric flux:

\begin{align}
L_{\star} &= L_{\odot} \, 10^{0.4\big[M_{\rm bol, \odot} - \left(M_{\rm V} + {\rm BC}\right)\big]},
\end{align}

\noindent where ${\rm BC}$ is the bolometric correction applied to the absolute magnitude of the RR Lyrae variable in the Johnson V band; see \cite{sandage1990} for a table of RR Lyrae bolometric corrections. The luminosity distance can then be
computed with an observed (mean) brightness $b_{\star}$. However, absolute magnitudes can only be inferred from the period in infrared bands \citep{catelan2004}; consequently, we must rely upon other information to determine a particular variable’s absolute magnitude when considering visible wavelength data. \cite{chaboyer1999} provides a
relation between the absolute magnitude of RRab variables and the metallicity ([Fe/H]):

\begin{align}
\label{eq:absolute_magnitude} M_{\rm V} &= (0.23 \pm 0.04)\, [{\rm Fe/H}] + (0.93\pm 0.12).
\end{align}

RR Lyrae variables belong to one of three classes based on the shape of their light curve \citep{smith1995}; RRab variables are the most common, and we restrict our catalogue of RR Lyrae variables to those of type RRab. The terms
RR and RRab Lyrae variable stars will be used interchangeably throughout this paper.

The distance modulus ${\mu_{\rm RRL} \equiv m_{\rm V} - M_{\rm V} - A_{\rm V}}$, where $m_{\rm V}$, $M_{\rm V}$ are the apparent and absolute magnitudes in the Johnson V band, and $A_{\rm V}$ is the associated extinction, i.e. reddening due to dust and other intermediate material, provides a well-established logarithmic scale for stellar distances:

\begin{align}
\label{eq:distmod} \mu_{\rm RRL} &= 10 + 5 \Big(\log_{10} \, {d_{\rm RRL} \over[1 \, {\rm kpc}]}\Big).
\end{align}

The normalized RR Lyrae distance modulus uncertainty can be determined using error propagation:

\begin{align}
\label{eq:uncertainty} \delta \mu_{\rm RRL} &= \sqrt{\delta m_{\rm V}^{2} + \delta M_{\rm V}^{2} + A_{\rm V}^{2}}, \\
\label{eq:distmod_uncertainty} \epsilon_{\mu_{\rm RRL}} &\equiv {1 \over \mu_{\rm RRL}} \, \delta \mu_{\rm RRL},
\end{align}

\noindent where the uncertainty of a measured value $x$ is denoted $\delta x$.
Equation \eqref{eq:uncertainty} establishes the relationship
between the measured distance modulus uncertainty and absolute magnitude uncertainty. We proceed with using the distance modulus as a proxy for physical distance, as the associated uncertainty is typically an order-of-magnitude smaller. In order to ensure that we avoid identifying specious structures at large distance moduli, we set an upper limit on substructure size in physical space and then translate to an appropriate distance moduli spread $\Delta \mu_{\rm RRL}$.

\begin{figure*}
    \centering
    \includegraphics[width=0.9\linewidth]{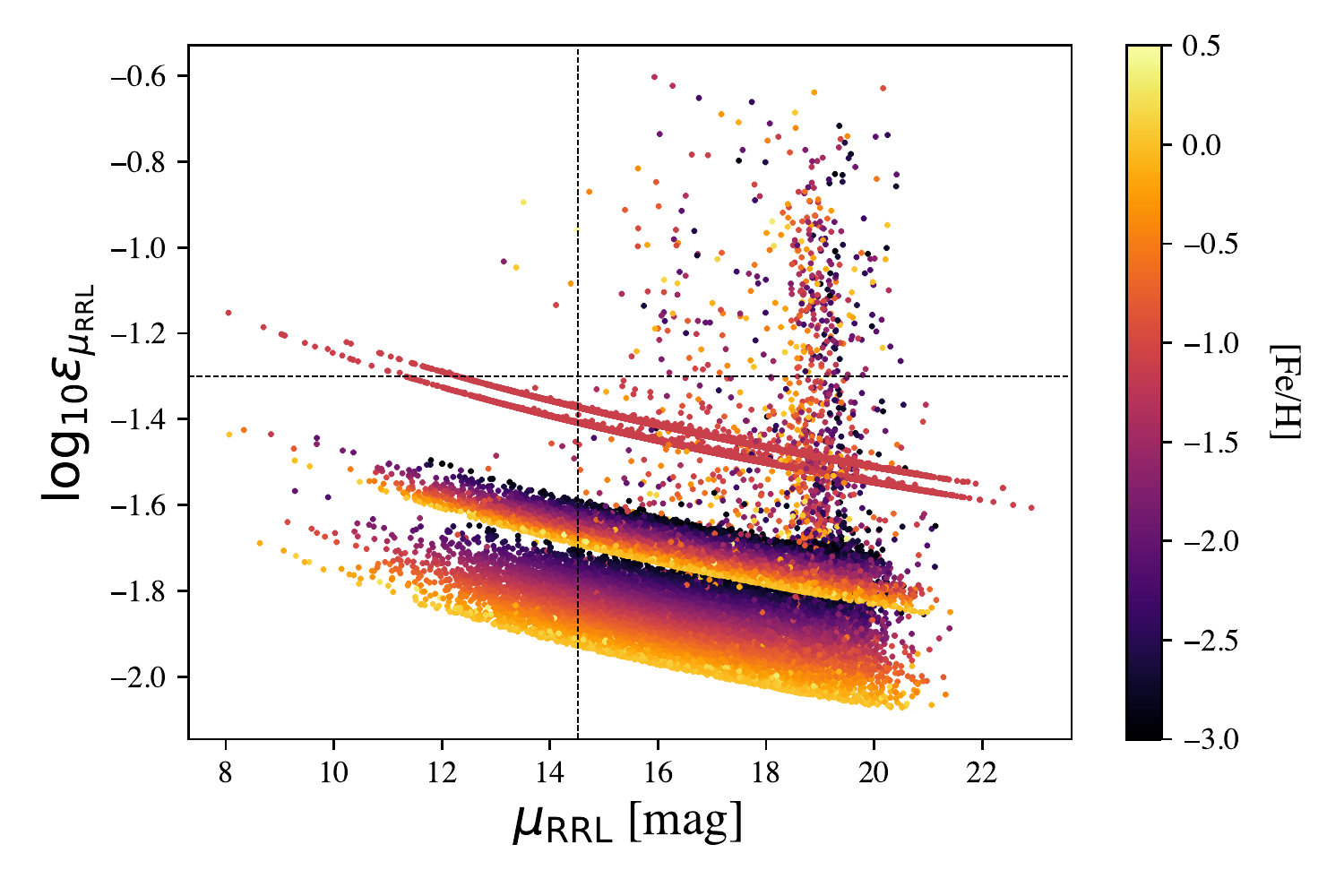}
    \caption{The distance modulus uncertainty $\epsilon_{\mu_{\rm RRL}}$ and distance modulus $\mu_{\rm RRL}$ of each RRab variable, where each data point is colour-coded by its reported or estimated metallicity. Dashed lines show the galactocentric distance (8 kpc) translated into distance modulus units, as well as the distance modulus uncertainty threshold consistent with $5\%$ of the computed distance modulus. A large number of RR Lyrae variables in the data set have a reported {\it Gaia} $G$-band value $G \simeq 19$, which explains the collection of variables with a distance modulus $\mu_{\rm RRL} \simeq 19$ that appears independent of the major $\log_{10}\epsilon_{\mu_{\rm RRL}}$ trends.}
    \label{fig:eps_distmod_vs_distmod}
\end{figure*}

\subsection{RRab variable data from the {\it Gaia} mission}

RR Lyrae variable stars have been identified using the {\it Gaia} mission's Specific Object Study (SOS) pipeline \citep{clementini2019}. The SOS pipeline ingested DR2 time-series photometry from the second data release (DR2) in the {\it Gaia} multi-bands ($G$, $G_{BP}$, and $G_{RP}$). Processing was predicated upon the period-amplitude and period-luminosity relations found within each variable star's $G$-band light curve, as well as colour-magnitude relations reported in DR2. Each variable star's best classification was then reported as part of DR2; for this study, we retained variables of type RRab for further analysis.

The translation between the Johnson V-band and {\it Gaia} multi-bands can be approximated in the following way:

\begin{align}
m_{\rm V} &= G - \sum_{n=0}^{2} a_{n}(G_{BP} - G_{RP})^{n}, \\
\delta m_{\rm V} &= \sqrt{\delta G^{2} + (a_{1}+2a_{2}(G_{BP}-G_{RP}))^{2} (\delta G_{BP}^{2} + \delta G_{RP}^{2}}),
\end{align}

\noindent where $\{a_{n}\}$ are polynomial coefficients fit to {\it Gaia} DR2 data\footnote{\url{https://gea.esac.esa.int/archive/documentation/GDR2/Data_processing/chap_cu5pho/sec_cu5pho_calibr/ssec_cu5pho_PhotTransf.html}}. In cases where the ${(G_{BP} - G_{RP})}$ colours or blue/red-band uncertainties were unavailable in the {\it Gaia} DR2 catalogue, we used, if available in the {\it Gaia} EDR3 catalogue, error propagation from the mean fluxes in each band with their associated uncertainties to compute ${G_{BP} - G_{RP}}$, $\delta G_{BP}$, and $\delta G_{RP}$. The reported (or inferred) {\it Gaia} magnitudes, and their uncertainties, could be used to compute ${(m_{\rm V}, \delta m_{\rm V})}$. If there was insufficient data available to make such inferences, the RRab variable datum was discarded.

The absolute magnitude and its associated uncertainty can be determined using Equation \eqref{eq:absolute_magnitude}:

\begin{align}
{\delta M_{\rm V} =  0.23 \, (\delta[{\rm Fe/H}]) + 0.04 \, \big| [{\rm Fe/H}] \big| + 0.12},
\end{align}

\noindent where $\delta[{\rm Fe/H}]$ is the reported metallicity uncertainty. RRab variables without a known metallicity were assigned the median value of the well-defined RRab variable sample, with an associated uncertainty equal to half the total metallicity domain, ${\delta[{\rm Fe/H}] = {1\over 2}[\max([{\rm Fe/H]})- \min([{\rm Fe/H]})]}$.

The extinction due to intermediate dust was computed using a publicly available data cube containing a 3D map of Milky Way dust; see \cite{green2018, green2019} for more details, as well as \cite{schlafly2011} to see how reddening translates to extinction. This mapping provides an extinction vector along each sightline given as input, where each vector element symbolizes the extinction value $A_{\rm V}(\mu)$ at that particular distance modulus along the sightline. Each ($A_{\rm V}$, $\delta A_{\rm V}$) pair was computed using the following scheme for each RR Lyrae variable star; if the distance modulus was within the range of support, the median and standard deviation of the five nearest vector elements were collected. In the event the unreddened distance modulus of the star was beyond the reported range of support of a well-defined extinction vector, the extinction value took on the maximum (far-field) vector value and the uncertainty was left equal to zero. This choice was motivated by Figures 1-5 in \citep{green2019}, where it is shown that the total extinction integrated to infinity is of the same order as the integrated extinction in the domain of the extinction vector $\mu \in (0, \mu_{\rm max}\sim 15]$, i.e.

\begin{align}
\int_{0}^{\infty} {\dd A_{\rm V} \over \dd \mu} \, \dd \mu \simeq \int_{0}^{\mu_{\rm max}} {\dd A_{\rm V} \over \dd \mu} \, \dd \mu.
\end{align}

RR Lyrae variable stars whose corresponding extinction vector contained ${\tt NaN}$ values were assigned the following ordered pair values: ${(\tilde{A}_{\rm V}, {1\over 2}[\max(A_{\rm V}) - \min(A_{\rm V})]})$, where $\tilde{x}$ represents the median of a collection of $x$ values.

In order to render out false positive stellar stream identifications from the hierarchical clustering trees we construct later on (see \S \ref{subsec:hierarchical_clustering}), we turn to the {\it Gaia} EDR3 catalogue \citep{gaia2020, lindegren2020} for RR Lyrae variable star proper motions on the sky plane. The newest data release from the {\it Gaia} mission contains full astrometry for $\sim 1.5$ billion sources, including parallaxes and proper motions. The {\it Gaia} cross-matching step is achieved using an ADQL query\footnote{\url{https://gea.esac.esa.int/archive/}}, in which matches are returned if the source ID is in agreement between the DR2 and EDR3 catalogs. There are few radial velocities available from the {\it Gaia} DR2 catalogue available for these stars, so we restrict ourselves to only considering motion on the sky plane. We retained 91234 RRab variables from the {\it Gaia} DR2 variable catalogue after the entire data collection and refinement routine, discarding data with missing or ill-defined values critical to the computation of the distance modulus. Figure \ref{fig:eps_distmod_vs_distmod} shows that the majority of the RRab variables have an inferred distance modulus uncertainty $\lesssim 5\%$ of the distance modulus itself, thus providing support for the usage of RR Lyrae variables in finding Milky Way substructures.

\begin{figure*}
    \centering
    \includegraphics[width=\linewidth]{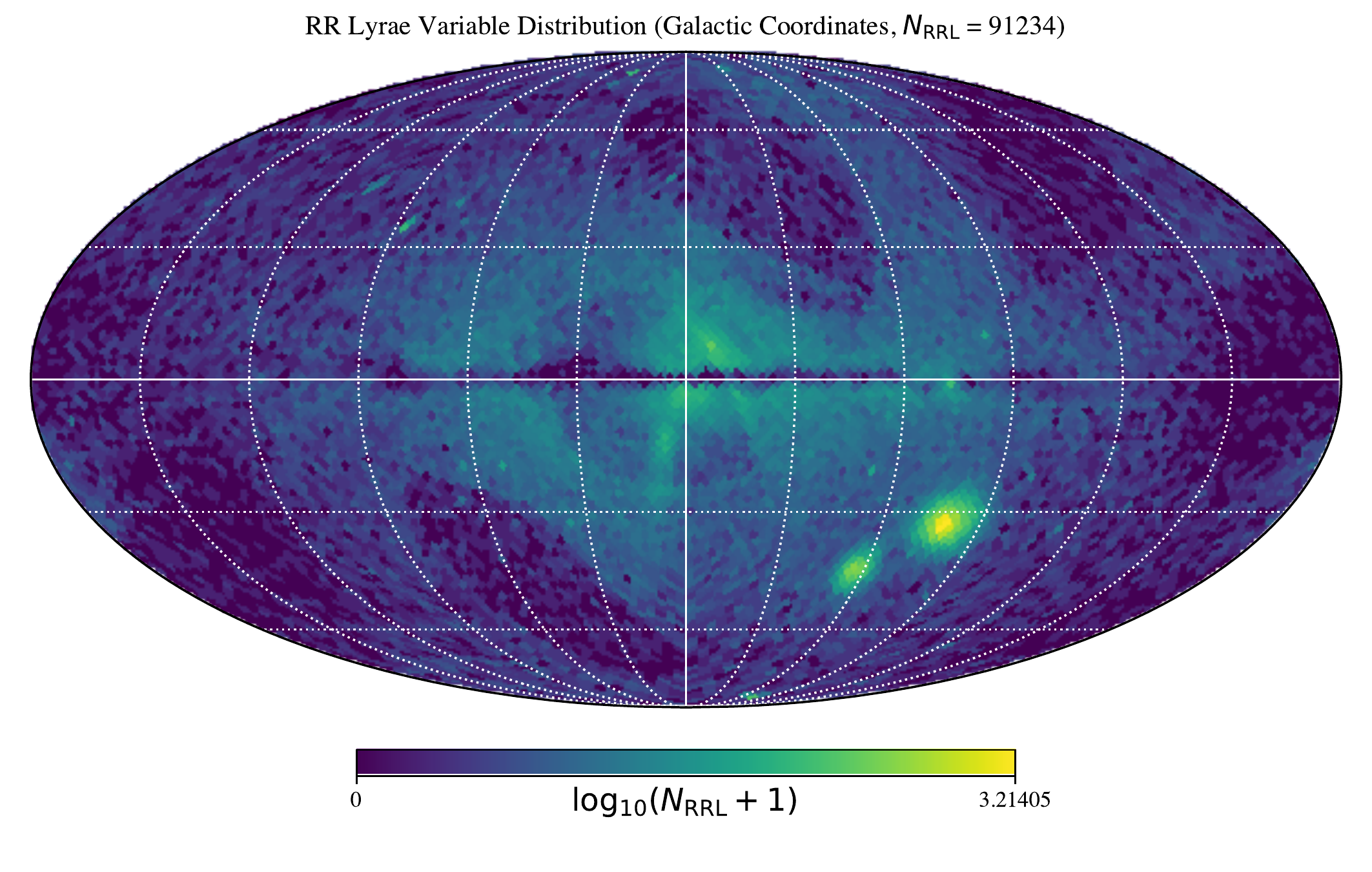}
    \caption{A HEALPix number density map ({$N_{\rm pixels} = 12288$}, see \S \ref{sec:results}) of the RR Lyrae variables (type RRab) from the {\it Gaia} DR2 catalogue retained for this study. Notable regions with an overdensity of RR Lyrae variables include the Galactic center and Magellanic Clouds. The projection is in Galactic coordinates $(\ell, b)$, where the grid is in thirty degree intervals.} 
    \label{fig:RRL_scatterplot}
\end{figure*}

\subsection{Particle-based, agglomerative, hierarchical clustering}
\label{subsec:hierarchical_clustering}

A hierarchical clustering algorithm needs an $\mathbb{R}^{n\times n}$ matrix to encode the separation distances between the $n$ objects that will populate the tree. Choosing the metric with which to calculate the distances depends on the nature of the
objects; the standard metric is Euclidean, although distances in phase space and nonphysical spaces that incorporate
stellar attributes such as metallicity have been used to identify potential clusters in a variety of astronomical
contexts \citep[e.g.,][]{maciejewski2009, desilva2015}. Using the catalogue data, we can construct three-dimensional vectors ${{\bf x} \equiv (x,y,z)}$ to get the distance between two
variable stars with the Euclidean metric:

\begin{align}
\label{eq:euclidean_distance} D({\bf x}, {\bf x}') &= \sqrt{(x-x')^{2} + (y-y')^{2} + (z-z')^{2}}, \\
x &\equiv \mu_{\rm RRL} \cos \alpha \cos \delta, \\
y &\equiv \mu_{\rm RRL} \sin \alpha \cos \delta, \\
z &\equiv \mu_{\rm RRL} \sin \delta.
\end{align}

\noindent where $\mu_{\rm RRL}$ is the distance modulus. Each RR Lyrae variable's angular
coordinates are expressed by its right ascension (${\phi = \alpha}$) and declination (${\theta = \delta}$). The distance between RR Lyrae variables in this metric have units of magnitude, which can be directly translated into a physical distance.

There are a number of choices in how to quantify the separation between two clusters, and for this analysis we have focused on the average linkage. The average linkage distance between clusters $p$ and $q$ is

\begin{align}
L(p,q) = {1\over n_{p}n_{q}} \sum_{i=1}^{n_{p}}\sum_{j=1}^{n_{q}} D({\bf x}_{p,i}, \, {\bf x}_{q,j}),
\end{align}

\noindent where $D({\bf x}_{p,i}, \, {\bf x}_{q,j})$ is the distance (as defined in Equation \eqref{eq:euclidean_distance}) between the $i$th and $j$th variables in clusters $p$ and $q$, respectively. This linkage method is a compromise between single linkage (minimum distance between points in
compared clusters, susceptible to imbalances and ``chaining”) and complete linkage (maximum distance between points
in compared clusters, tends to generate clusters that are roughly uniform in size and shape).

The algorithm proceeds as follows: if we have $n$ clusters ${c_{1}, \dots , c_{n}}$, the ${n(n - 1)/2}$ unique linkage distances are computed. Let us say that the minimum linkage distance is between clusters $c_{i}, c_{j}$. The sets of RR Lyrae variables
describing these two clusters are then combined into a single cluster $c_{ij}$, and at the next step there are ${n - 1}$
clusters: ${c_{1}, \dots, c_{ij}, \dots, c_{n-1}}$. Once there is only a single cluster remaining, the algorithm terminates.

Agglomerative clustering does not provide a straightforward method for determining the optimal set of clusters, as
the initial and final number of clusters are fixed. Intermediate steps can be analyzed by ``cutting” the tree constructed
from the condensed distance matrix at a specific height such that all identified clusters have an average linkage distance
greater than or equal to that height. \cite{langfelder2007} provides a general method of tree-cutting that determines
the best choice of clusters. If we cut the tree at a large number of heights and count the corresponding number of
identified clusters at each height, we can identify a particular configuration that is resistant to further clustering.
For example, if the minimum average linkage distance between $n$ clusters is $h_{0}$ and the same minimum distance
between $m - 1$ clusters is $h_{1}$, it stands to reason that the variables are neatly classified into $n$ clusters if ${|h_{1} - h_{0}| \gg \Delta h}$, where $\Delta h$ is a tree-cutting height interval. Quantifying this height separation can be done by setting up an array of heights at which the tree is cut (separated by $\Delta h$), populating a corresponding array with the number of branches (clusters) at each height, and then computing the mode of that array to determine
the optimal number of clusters. However, this procedure would then obscure interesting phenomena at the scales of globular clusters and stellar streams.

With agglomerative clustering, there is a suitable alternative; we can determine scales at which there are comparatively many clusters populated by a statistically significant number of RR Lyrae variables. Using the tree-cutting process, we compute the mean and standard deviation of cluster populations at each height. The heights at which the tree are cut are evenly spaced in log space between ${h_{\rm min} = 0.1 \, {\rm mag}}$ and ${h_{\rm max} = 10 \, {\rm mag}}$; the
motivation for this scheme is that we are interested in a deeper analysis of the stellar halo's substructure than in
structures at Galactic scales.

\begin{figure}
    \centering
    \includegraphics[width=\linewidth]{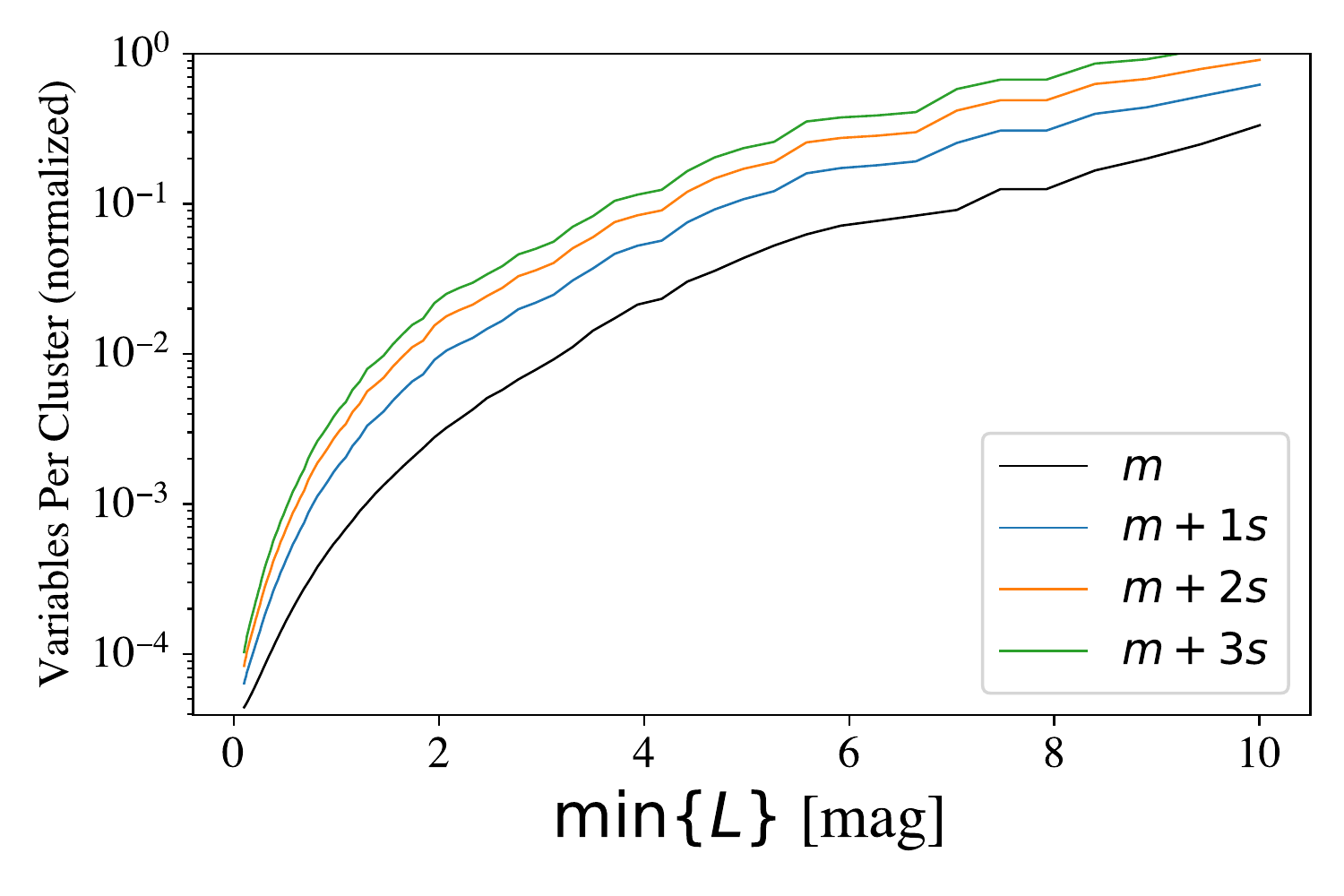}
    \caption{The mean number of RR Lyrae variables per cluster as a function of clustering scale, where the requisite hierarchical clustering tree was constructed from the catalogue subsample displayed in Figure \ref{fig:clusters_at_different_scales}). Each cluster contains, on average, $m$ RR Lyrae variables with a standard deviation $s$ at each of the listed clustering scales. Clusters whose variable population exceeded ${m + 3s}$ were regarded as statistically significant.}
    \label{fig:means_and_stdevs}
\end{figure}

Figure \ref{fig:means_and_stdevs} shows that at intermediate distance scales, the criterion for being categorized as a significantly populated cluster becomes more
stringent. When the variables first begin to agglomerate (${\min \{L \}\sim 0.1 \, {\rm mag}}$, where $\{L\}$ is the set of linkage distances), there may be two or three variable stars
that are considerably close; the mean number of variables per cluster is very close to 1 and the standard deviation
is small in this regime, so this would be regarded as a cluster of interest. Such clusters should be approached with
skepticism, given that the computed distance uncertainties are larger than these separation distances. On the largest
scales (${\min \{L\} \gtrsim 10}$), the standard deviation is considerably higher than the mean, so a cluster with a variable
population $\approx N_{\rm RRL, total}$ would be the only significant cluster. If there are only two or three clusters remaining, however, then
this would yield a high significant-to-insignificant-cluster ratio.

\begin{figure*}
    \centering
    \includegraphics[width=\linewidth]{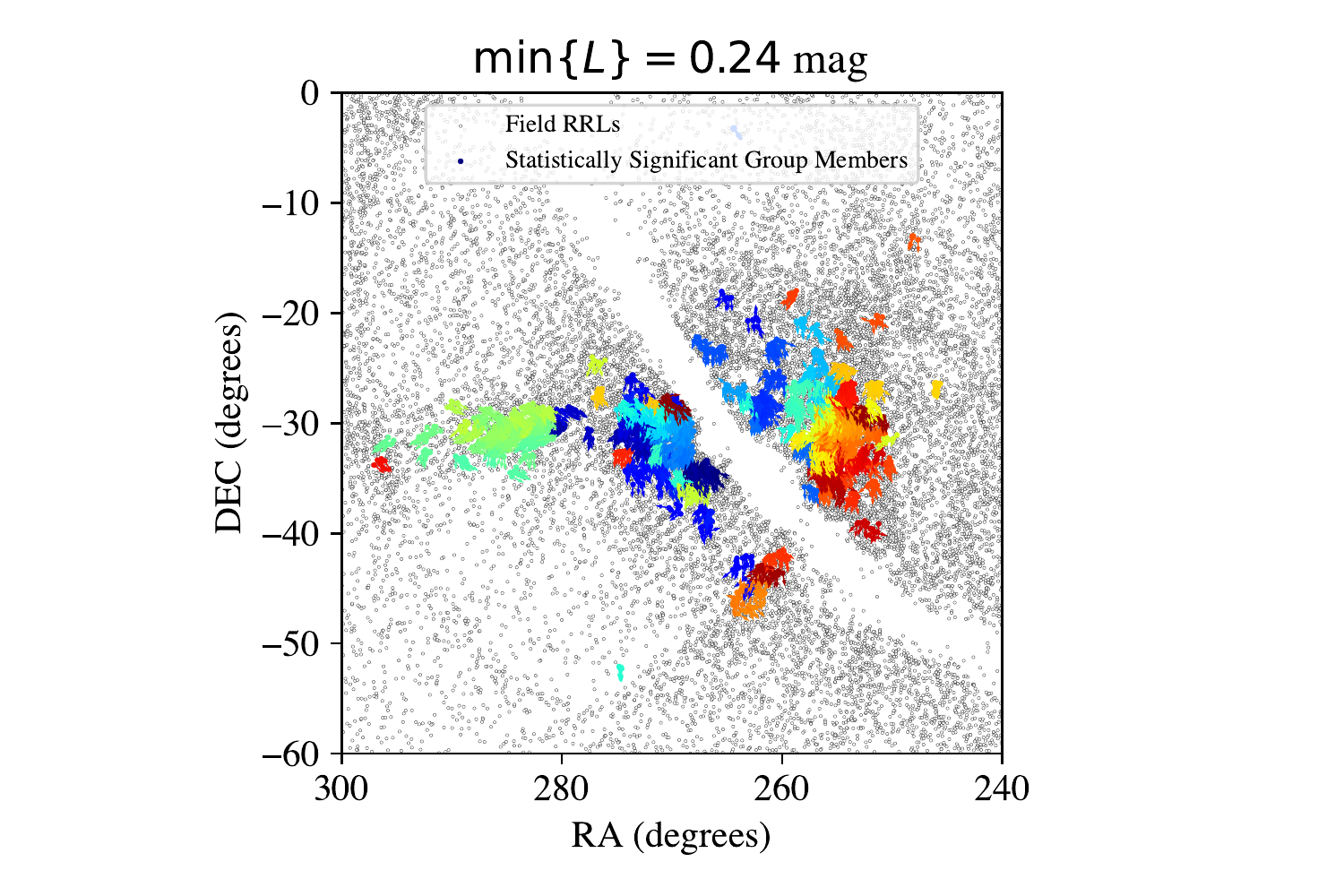}
    \caption{Hierarchical clustering at a scale (${\min \{L\} \simeq 0.24 \, {\rm mag}}$) that contains many RR Lyrae variables belonging to statistically significant groupings with correlated proper motions (${\tilde{X}_{\theta} \geq 0.98}$, see Equation \eqref{eq:setDef}). Variables colour-coded by their identified grouping are indicated by bigger markers to distinguish from field variables at this scale, and (normalized) arrows indicate {\it Gaia} proper motion directions. The hierarchical clustering tree used to generate this figure contains all RR Lyrae variables in the displayed RA/DEC window, where subsequent figures use HEALPix subregions. The displayed region includes both the Galactic bulge (mostly blue and red groupings) and the Sagittarius Dwarf Spheroidal Galaxy (mostly green groupings), both of which contain many RR Lyrae variables and are, as a result, suitable for this illustration. Some of the groupings shown could be omitted from later figures, e.g. if their collective distance modulus spread indicates a separation distance in physical space of greater than $R_{\rm max} = 1 \, {\rm kpc}$.}
    \label{fig:clusters_at_different_scales}
\end{figure*}

We collected all of the statistically significant clusters (i.e., ${N_{\star} \geq m + 3s}$, where $m$ is the mean number of RR Lyrae variables per cluster and $s$ is the associated standard deviation) at clustering scales ($N_{\rm cuts} = 80$ evenly separated in log space between $h_{\rm min}$ and $h_{\rm max}$ defined above) where the ratio of statistically significant to total clusters was maximized. Figure \ref{fig:clusters_at_different_scales} shows the results of clustering stars within a subregion of the sky plane, where stars belonging to statistically significant groupings are colour-coded by a grouping ID at a small clustering scale; we should expect to find groupings of RR Lyrae variables consistent with stellar streams at a variety of clustering scales, and the logarithmic spacing of our cuts ensures sampling of the trees' variety of branch sizes with a preference towards clustering scales where $\min\{L\} \lesssim 2 \, {\rm mag}$. Collecting every cluster identified across all significant clustering scales would admit redundant groupings; from these redundancies, one was retained from each for further analysis while potential redundancies are listed in the Appendix (\S \ref{sec:potentially_extraneous_candidates}).

The relevance of the identified groupings can be tested using {\it Gaia} EDR3 proper motions. Stellar streams moving on non-chaotic orbits through the Galactic tidal field will have well-correlated configurations in velocity space. This instructs our usage of velocity information in the form of sky plane proper motions. To start, we determine each grouping's median proper motion direction $\hat{v}$, written as a two-dimensional vector on the sky plane:

\begin{align}
\hat{v} &\equiv {\langle \tilde{\mu}_{\alpha}^{*}, \tilde{\mu}_{\delta} \rangle\over \sqrt{\tilde{\mu}_{\alpha}^{*^2} + \tilde{\mu}_{\delta}^{2}}},
\end{align}

\noindent where ${\mu_{\alpha}^{*} \equiv \mu_{\alpha}\cos\delta}$. These proper motion components are not to be confused with the distance modulus $\mu_{\rm RRL}$. We can determine the alignment of the $i$th star of the grouping by taking the dot product between its proper motion and the grouping's median proper motion:

\begin{align}
\label{eq:dotProduct} \cos \theta_{i} &=  \hat{v} \cdot \hat{v}_{i} = {(\tilde{\mu}_{\alpha}^{*} \, \mu_{{\alpha},i}^{*} + \tilde{\mu}_{\delta}\, \mu_{{\delta},i})\over \sqrt{\left(\tilde{\mu}_{\alpha}^{* ^ 2} + \tilde{\mu}_{\delta}^{2}\right)\left(\mu_{{\alpha},i}^{* ^ 2} + \mu_{{\delta},i}^{2}\right)}}.
\end{align}

The set 

\begin{align}
\label{eq:setDef} X_{\theta}\equiv \{\cos\theta_{1}, \dots, \cos\theta_{N}\},
\end{align}

\noindent which is a measure of how well each of the $N$ stellar proper motions within the grouping are aligned with the median proper motion, will thus contain values ranging from -1 (antiparallel) to 1 (parallel). If the median of this set is below ${\tilde{X}_{\theta, {\rm min}} = 0.98}$, the grouping is discarded. 

The proper motion validation layer is dependent on the following assumptions: the median proper motion vector is appropriate for comparison with all grouping members, and that the data within a grouping is only mildly heteroscedastic, i.e., that the statistical deviations from the median proper motion do not vary from one part of the grouping to the next. The first assumption breaks down for groupings distributed across large regions of the sky plane, and the second assumption breaks down for systems that have been disturbed by external objects like the Galactic Bar. The Pal 5 stellar stream has both of these qualities \citep{pearson2017}, which explains why our routine can only identify the stream's core (see \S \ref{subsec:pal5_gaia}). 

We only retain groupings with parameters similar to \cite{sesar2013}:
${15 \leq N_{\rm RRL} \leq 40}$ and ${R_{\rm grouping} \leq R_{\rm max}= 1 \, {\rm kpc}}$; see Appendix \ref{appendix:amuse} for motivation of the chosen $\tilde{X}_{\theta, {\rm min}}$, $R_{\rm max}$ values. The grouping size $R$ translates to a (positive) distance modulus spread $\Delta \mu$ that is dependent on how far away the grouping is from the observer (i.e., distance modulus $\mu$, Equation \eqref{eq:distmod}):

\begin{align}
\label{eq:r_from_distmod} {R \over [1 \, {\rm kpc}]} &= 10^{(\mu + \Delta \mu / 2)/5 - 2} - 10^{(\mu - \Delta \mu / 2)/5 - 2}, \\
\label{eq:distmod_spread} \Delta \mu (\mu, R=R_{\rm max}) &= 10 \, \log_{10} \Bigg({1\over 2}\bigg[10^{2-\mu/5} + \sqrt{10^{2(2-\mu/5)}+4}\bigg]\Bigg).
\end{align}

We collect all potentially relevant groupings, regardless of size, but then discard potentially spurious groupings whose radius in distance modulus units indicates a physical size greater than ${R=R_{\rm max}}$.

Once the RR Lyrae streams candidates were properly identified, we applied an ellipsoid fitting algorithm \citep{bazhin2019} such that the principal axis vectors $\{{\bf e}_{k}\}$ of the best-fit ellipsoid are provided. Given our interest
in globular clusters and stellar streams, we applied the following scheme to the set of principal axis lengths $\{e_{k}\}$: if
the eccentricity ${e \equiv
\sqrt{
1 - {
\min(\{e_{k}\})/
\max(\{e_{k}\})}}}$ was $\leq 0.2$, the grouping
was classified as a globular cluster-like grouping (see \cite{harris1979, staneva1996} for justification),
and if $e \geq 0.7$ it was identified as a stellar stream-like grouping \citep{martin2010}. Groupings with eccentricities
in the intermediate range could in principle be admitted, but classified with the designation ``other”; in all cases, the grouping radius is
defined as ${R_{\rm grouping} \equiv \max \{e_{k}\}}$.

Most of the identified groupings are classified as stellar streams within this paradigm. If there are outliers (which are permissible but unlikely with average linkage) or the number of grouped RR Lyraes is small, this eccentricity calculation would be susceptible to unacceptably high variations if new RR Lyraes were included. In such cases, this eccentricity computation should be thought of as a first-order estimate. It is beyond the scope of this study, but in instances where the internal structure of a grouping is important, ellipsoid fitting is ill-advised. A minimum spanning tree \citep{kruskal1956}, for example, could be used to further analyze the integrity of the grouping instead.

\subsection{Finding Palomar 5 with {\it Gaia} RR Lyrae variables}
\label{subsec:pal5_gaia}
 
A collection of RR Lyraes from the {\it Gaia} DR2 and Pan-STARRS1 \citep{sesar2017} catalogues were recently used to analyze the kinematic properties of the Palomar 5 stellar stream \citep{price-whelan2019}, a system whose properties make it suitable for analyses of the Galactic potential \citep[e.g.,][]{bovy2016, starkman2020}. The publicly available dataset included the 3300 RR Lyrae variables found in the appropriate section of the sky plane, but in our pre-processing we only retained the 2348 variables of type RRab. A set of stellar stream model parameters were inferred with Markov chain Monte Carlo posterior sampling, and each of the catalogue variables were given a posterior membership probability.

\begin{figure}
\centering
\includegraphics[width=\linewidth]{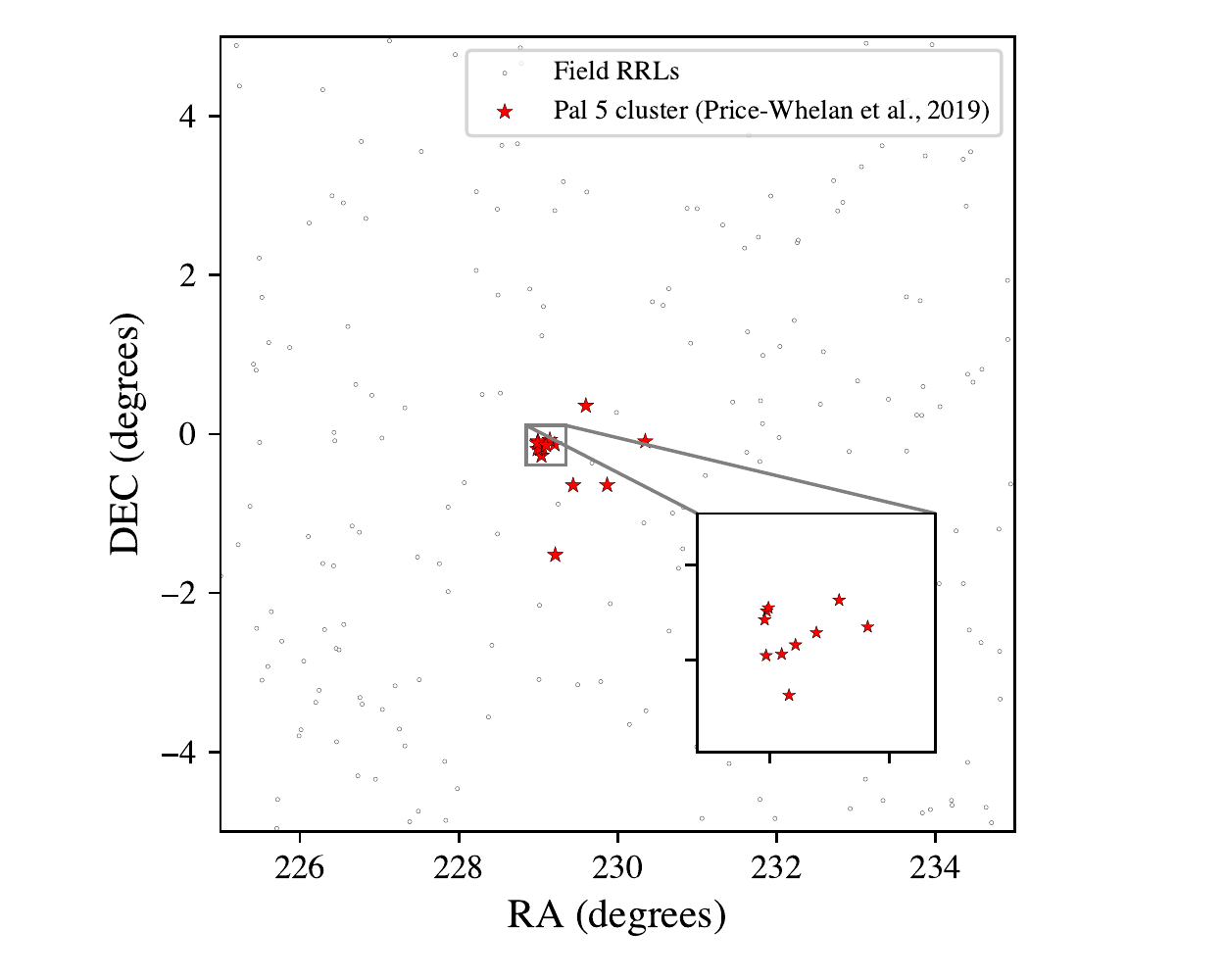}
\caption{An RR Lyrae grouping identified by the clustering algorithm described in \S \ref{subsec:hierarchical_clustering} that is consistent with the Palomar 5 globular cluster. In comparing with a map of nearby RR Lyrae variables colour-coded by membership probability \citep[][Figure 3]{price-whelan2019}, we see that the inset RR Lyrae variables are within Pal 5's Jacobi radius (see Appendix \ref{appendix:amuse} for a definition).}
\label{fig:Pal5_cluster}
\end{figure}

Figure \ref{fig:Pal5_cluster} shows the one grouping identified in the refined {\it Gaia} catalogue via hierarchical clustering, and it is consistent with the Palomar 5 globular cluster. This grouping contained fifteen RRab variables, eight of which were within the globular cluster's reported Jacobi radius. We used the reported physical distances (in kiloparsecs) to infer each variable's distance modulus via Equation \eqref{eq:distmod}. While the grouping has an eccentricity ${e_{\rm grouping} \simeq 0.9}$, which qualifies as a stellar stream in our paradigm, this is attributable to the external variables included in the grouping and the general nature of average linkage. The grouping has a median distance modulus of $\mu_{\rm grouping} \simeq 16.6 \, {\rm mag}$, which is roughly consistent with Palomar 5. Thirteen of the fifteen variables identified by our algorithm were awarded an MCMC membership probability of ${P\gtrsim 0.5}$; this, along with the absence of false positives, provides a reliable cross-check of our cluster identification routine against a known substructure.

\section{Results}
\label{sec:results}

In order to find archaeologically relevant substructures from the {\it Gaia} data via hierarchical clustering, we must make an informed decision of how to allocate the variables appropriately (e.g., consider employing random subsampling \citep{khoperskov2020} or breaking up the catalogue into smaller trees.) If there is a substructure of $N_{\star}$ RR Lyrae variables found in a catalogue of size $N_{\rm catalogue}$ (assuming the described grouping cultivation routine is guaranteed to find it), then the probability that it is found within a randomly selected subset of size $N_{\rm subset}$ is

\begin{align}
P(N_{\star}) &\simeq {N_{\rm subset}! \over N_{\rm catalogue}!} {(N_{\rm catalogue}-N_{\star})! \over (N_{\rm subset} - N_{\star})!}.
\end{align}

The detection probability $P(N_{\star})$ decreases dramatically as $N_{\rm subset}$ decreases, which is a quantitative motivation for choosing as large a catalogue of RR Lyrae variables as possible. However, the tree cutting process is not guaranteed to be well-suited for an all sky plane catalogue; indeed, the results shown in Figures \ref{fig:clusters_at_different_scales} and \ref{fig:Pal5_cluster} were determined using hierarchical clustering trees comprised of RR Lyrae variables from a comparatively small patch of the sky plane. 

These conflicting attributes of hierarchical clustering trees motivate the following choice: we created forty-eight hierarchical clustering trees covering distinct HEALPix\footnote{\url{https://healpix.jpl.nasa.gov/}} subregions of equal sizes (${A_{\rm subregion} \simeq 859 \, {\rm deg}^{2}}$) on the sky plane, following the RING ordering convention. The union of each tree's subpopulations is the entire catalogue of 91234 variables shown in Figure \ref{fig:RRL_scatterplot}. We iterate through each of the forty-eight trees, collecting substructures with the desired proper motion and physical size attributes into a database.

\begin{table*}
    \centering
    \caption{Bulk attributes (e.g., median declination) of RR Lyrae substructures identified using the methods delineated in \S \ref{sec:methods} and displayed in Figure \ref{fig:substructure_scatter}, sorted by distance modulus. The reported radius is converted into physical units using Equation \eqref{eq:r_from_distmod}. The eccentricities of the identified substructures suggest pronounced elongation consistent with stellar streams. Substructures whose distance modulus separation from the nearest Milky Way globular cluster \citep{harris2010} is less than $\Delta \mu(\tilde{\mu}, R=R_{\rm max})$ (see Equation \eqref{eq:distmod_spread}) are more likely to be associated with said globular clusters, while the others warrant further investigation.}
    \label{tab:rrl_substructures}
\begin{tabular}{c|c|c|c|c|c|c|c|c|c}
\hline
ID &  $\tilde{\mu}$ [mag] &          $\tilde{\alpha}$ [deg] &        $\tilde{\delta}$ [deg] &    $R$ [pc] &  $N_{\star}$ &  Eccentricity & Nearest GC & $d_{{\rm substruct} \to {\rm GC}}$ [mag] & $\Delta \mu (\tilde{\mu}, R=1 \, {\rm kpc})$ [mag] \\
\hline
\hline
0  &             12.64 &  245.9 & -26.5 &  225.11 &      29 &      0.996877 & NGC 6121 & 0.93 & 0.64 \\
3  &             13.81 &  201.6 & -47.4 &  749.46 &      16 &      0.983881 & NGC 5139 & 0.23 & 0.38 \\
9  &             14.25 &  154.4 & -46.4 &  401.87 &      22 &      0.998333 & NGC 3201 & 0.80 & 0.31 \\
13 &             14.39 &  229.6 &   2.1 &  818.17 &      27 &      0.998918 & NGC 5904 & 0.02 & 0.29 \\
15 &             14.58 &  263.0 & -67.1 &  922.35 &      22 &      0.915398 & NGC 6362 & 0.18 & 0.26\\
16 &             14.78 &  248.1 & -13.1 &  160.32 &      20 &      0.967803 & NGC 6171 & 0.75 & 0.24 \\
17 &             15.00 &  189.9 & -26.7 &  484.83 &      18 &      0.968888 & NGC 4590 & 0.06 & 0.22 \\
19 &             15.13 &  255.3 & -30.1 &  274.61 &      26 &      0.942493 & NGC 6266 & 0.97 & 0.20 \\
22 &             15.39 &   78.5 & -40.1 &  228.64 &      15 &      0.997036 & NGC 1851 & 0.03 & 0.18 \\
24 &             15.62 &  255.3 & -30.1 &  454.92 &      20 &      0.997226 & NGC 6316 & 1.17 & 0.16 \\
25 &             16.30 &  313.4 & -12.5 &  526.09 &      26 &      0.995423 & NGC 6981 & 0.15 & 0.12 \\
27 &             16.32 &  308.6 &   7.4 &  382.78 &      26 &      0.981308 & NGC 6934 & 0.35 & 0.12 \\
28 &             16.58 &  264.4 &  -3.2 &  221.74 &      15 &      0.981297 & IC 1257 & 1.43 & 0.10 \\
29 &             17.03 &  225.1 & -82.2 &  737.69 &      21 &      0.998445 & IC 4499 & 0.66 & 0.09 \\
30 &             17.14 &  225.1 & -82.2 &  955.57 &      21 &      0.987165 & IC 4499 & 0.77 & 0.08 \\
31 &             18.71 &   96.8 & -70.1 &  741.74 &      15 &      0.937961 & E 3 & 5.73 & 0.04 \\
\hline
\end{tabular}
\end{table*}

\begin{figure*}
    \centering
    \includegraphics[width=\linewidth]{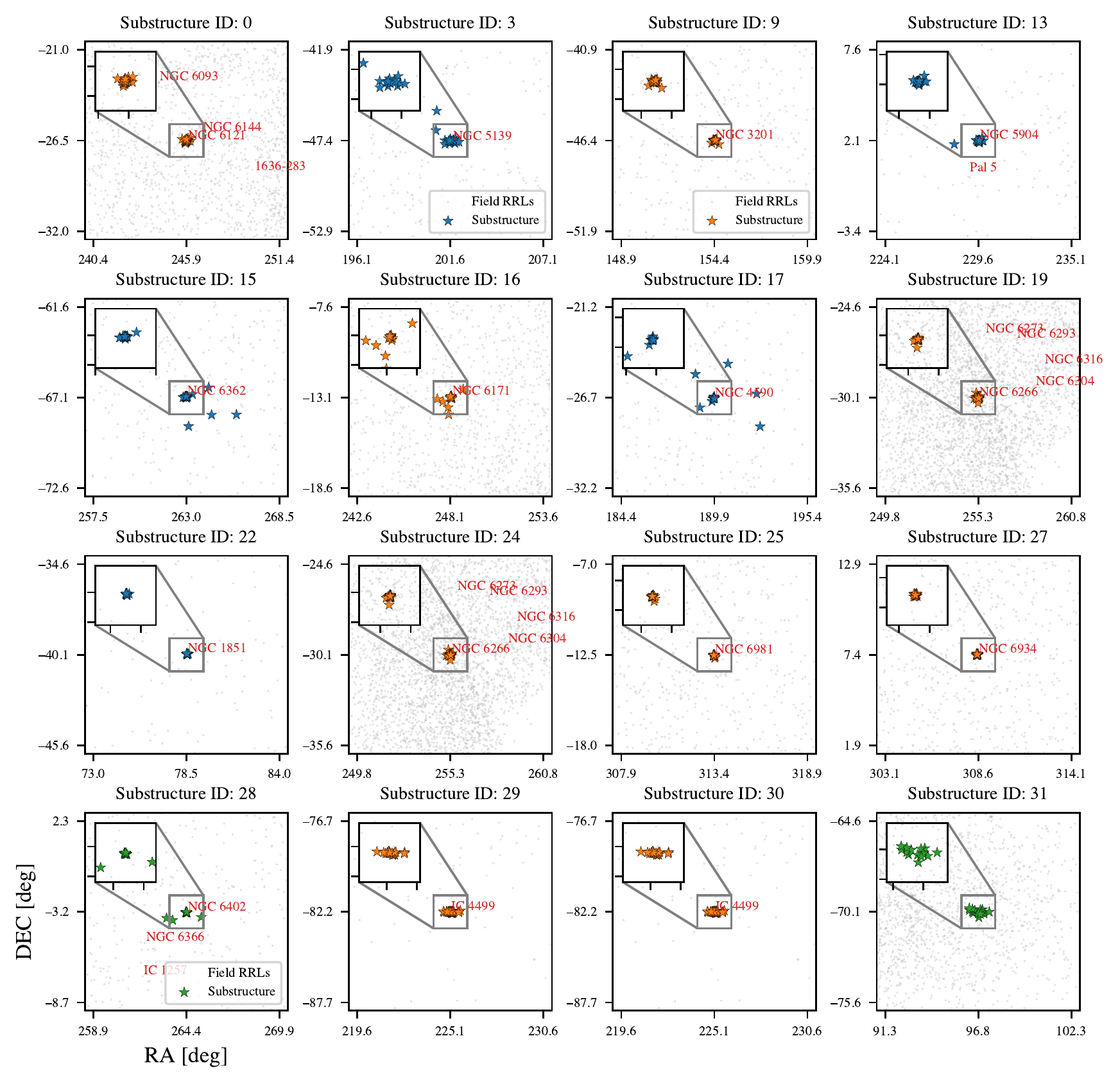}
    \caption{Substructures listed in Table \ref{tab:rrl_substructures}, colour-coded by substructure class as defined in \S \ref{sec:results}. Field RR Lyrae variables from the cultivated catalogue and local globular clusters on the sky plane are included. A ${2^{\circ} \times 2^{\circ}}$ inset in each panel provides a zoomed-in view of each of the substructure cores, most of which are consistent with Milky Way globular clusters.}
    \label{fig:substructure_scatter}
\end{figure*}

The resulting database contained thirty-two candidate substructures, broken up into Tables \ref{tab:rrl_substructures} and \ref{tab:rrl_substructures_appendix}, with a preference towards deeper analyses of the first group of substructures (shown in Figure \ref{fig:substructure_scatter}). The separation of substructure candidates took the following attributes into consideration: number of RR Lyrae variables, physical size, eccentricity, and proximity to the nearest globular cluster. If a substructure candidate had a larger physical size, it likely means that the constituent RR Lyrae variables are less likely to be associated with any parent globular cluster system. Substructures with IDs 2-4, for example, are RR Lyrae variable sets with, presumably, large intersections. Our stated preference for analyzing the substructures in Table \ref{tab:rrl_substructures} is, admittedly, a subjective one.

Upon inspection, there are three substructure classes in Table \ref{tab:rrl_substructures}: substructures within a kiloparsec of a Milky Way globular cluster (IDs 3, 13, 15, 17, 22), substructures whose connections to any particular globular cluster warrants further investigation (0, 9, 15, 16, 19, 24, 25, 27, 29, 30), and substructures that may be independent of the Milky Way globular cluster population (28, 31). Globular clusters with a first class substructure less than 1 kiloparsec away are highlighted in Table \ref{tab:GCs_of_interest}. Substructures 3 and 17 are likely related to the Fimbulthul \citep{ibata2019} and Fj\"{o}rm \citep{ibata2019b} streams, while substructures 13, 15, and 22 are in the vicinity of known globular cluster tidal tails \citep[e.g.,][]{piatti2020, ibata2021}.

The second class of substructures have less straightforward connections to neighboring globular clusters, and this is, in part, due to increasingly unforgiving distance modulus spreads. Substructures 0, 9, 15, 16, 25, and 27 are likely part of their listed nearest globular cluster systems, but the more distant substructure members causing high substructure eccentricities (or distance modulus uncertainties) likely caused an identified substructure center separation ${d_{\rm{substruct}\to {\rm GC}} \gtrsim \Delta\mu(\tilde{\mu}, R=1\,{\rm kpc})}$.  

Substructures 19 and 24 (${\tilde{\mu}_{19}=15.13, \tilde{\mu}_{24}=15.62}$) are closest to NGC 6266 ($\mu\simeq 14.2$) and NGC 6316 ($\mu\simeq 15.09$), respectively, despite appearing to be virtually identical in Figure \ref{fig:substructure_scatter}. The high number density of RR Lyrae variables in this region, as well as the number of nearby Milky Way globular clusters, makes the attribution of these substructures to any particular globular custer of stellar stream beyond the scope of this study. Substructures 29/30 are nearly redundant and consistent with the IC 4499 globular cluster, a GC with a robust RR Lyrae population of type ab \citep{ferraro1995} and large tidal radius \citep{walker2011}. While the sky plane configuration of substructures 29/30 are encouraging for making connections to IC 4499, their distance moduli (${\tilde{\mu}_{29} = 17.03, \tilde{\mu}_{30} = 17.14}$) are larger than available estimates for the distance to IC 4499 \citep[e.g., ${\mu\simeq16.5}$,][]{storm2004}.

Substructure 28 (${\tilde{\mu} = 16.58}$) is closest to IC 1257 \citep[$\mu\simeq17.0$,][]{harris1997}, a small globular cluster in the mid-halo region of the Milky Way. The presence of the NGC 6402 and NGC 6366 globular clusters, whose distance moduli are $\mu \simeq 15$, introduces confusion. This substructure's association with any of these three globular clusters is dubious, as the distances are inconsistent by several kiloparsecs. Additionally, none of these globular clusters have known tidal tails \citep{piatti2020, ibata2021}. Therefore, this substructure is an appealing candidate for future studies.

The substructure with the largest distance modulus, substructure 31 (${\tilde{\mu}_{31} = 18.71}$), is only separated from the Large Magellanic Cloud by a few degrees on the sky plane and not likely to be associated with the E 3 globular cluster (as suggested in Table \ref{tab:GCs_of_interest}). With a physical distance of ${d_{\odot \to {\rm substruct}} \simeq 55.2 \, {\rm kpc}}$, substructure 31 is very likely a member of the Magellanic system. There is evidence that the Magellanic system was previously a triplet of dwarf galaxies before the LMC accreted one of its siblings \citep[e.g.][]{armstrong2018, mucciarelli2021}, and two kinematically distinct globular cluster populations \citep{piatti2019} suggests that the Magellanic system is a good place to search for artifacts of the hierarchical galaxy formation process. A runaway star cluster east of the LMC was recently discovered in a series of recent deep imaging surveys \citep{piatti2019b, piatti2021} near substructure 31; the reported age of the star cluster (${0.89_{-0.10}^{+0.11} \, {\rm Gyr}}$), however, is inconsistent with a substructure of Population II stars. An analysis of tile 5-9 in the VMC survey's RR Lyrae variable catalogue \citep{cusano2021} is likely needed to confirm the existence of this substructure and its properties.

\section{Discussion}
\label{sec:discussion}

Fourteen RR Lyrae variable streams have been identified from the Catalina survey at $>3.5 \sigma$ confidence \citep{mateu2018}, several of which are located near Milky Way globular clusters. We do not claim that the groupings shown in Figure \ref{fig:substructure_scatter} rise to this level of precision, as there is evidence that spectroscopy is required to confirm Galactic stellar stream candidates even in the {\it Gaia} era \citep{jeanbaptiste2017}. Instead, we will proceed with a discussion of the identified substructures, and their local environments, knowing that further investigation is needed. Follow-up studies would not only need to confirm each candidate grouping's legitimacy (likely through chemical abundance matching and an exploration of the entire phase space), but also to determine the entire scope of the candidate substructure beyond its population of RR Lyrae variables.

We determined each candidate grouping's nearest GC neighbour in the $(x,y,z)$ space defined in \S \ref{subsec:hierarchical_clustering}; the results are shown in Table \ref{tab:GCs_of_interest}. Eight of the fifteen GCs listed in this table were accreted during merger events \citep{kruijssen2020} and seven have been noted in the literature as having extratidal features in their outermost regions (\cite{piatti2020, ibata2021} and references therein). Additionally, {\it Gaia} EDR3 RR Lyrae variables have escaped from two of the fifteen globular clusters (NGC 5904 and NGC 1851) according to a recent study of proper motions and colour-magnitude diagrams \citep{abbas2021}. If there are new connections to be made between Milky Way GCs and the galaxy's RR Lyrae population via substructure identification, we suggest focusing on the following substructures where there is no previously known tidal tail despite being a potential progenitor traced back to a Galactic merger event: 0, 25, 29, 30. The {\tt Via Machinae} algorithm \citep{shih2021}, for example, is an unsupervised machine learning routine designed to detect stellar streams, and would likely be a suitable method for carrying out such a follow-up study.

The nuclear star clusters that remain from four of these merger events were recently identified using chemo-kinematic information \citep{pfeffer2021}, and our routine identified RR Lyrae substructures near two of them: NGC 5139 ($\omega$ Centauri), reportedly the remnant of the {\it Gaia}-Enceladus merger, and NGC 6934, the reported remnant of the Helmi streams merger (although NGC 6934's status as the merger remnant is less certain). Categorizing accreted globular clusters by their parent merger events, and even compartmentalizing the merger events themselves, is an ongoing area of research. A recent study of 23 known Galactic stellar streams, for example, demonstrated that clustering in orbital phase space is a fruitful method of identifying their sources \citep{bonaca2021}. Many of these streams are extended for $\gtrsim 100^{\circ}$ on the sky; our routine's inability to identify structures of this kind further motivates our focus on RR Lyrae substructures that may be components of stream cores in the neighbourhood of Milky Way GCs.

\begin{table*}
    \centering
    \caption{Milky Way globular clusters \citep[with reported distance moduli and Milky Way coordinates,][]{harris2010} near an identified RR Lyrae substructure presented in \S \ref{sec:results}. Each GC's potential association with a progenitor merger event \citep{kruijssen2020}, known extratidal features \citep{piatti2020} and well-studied Galactic stellar streams \citep{bonaca2021}, if applicable, are provided. (G1 denotes a symmetric tidal tail, G2 a feature outside of the Jacobi radius that is not necessarily a tidal tail, G3 no signature of extended structure, and SF a long tidal tail discovered using the {\tt STREAMFINDER} algorithm on {\it Gaia} DR2/EDR3 data \citep{ibata2021}. This classification scheme is only applied to previously known substructures and is independent of the results presented in this paper.) Highlighted globular cluster IDs indicate an instance where an RR Lyrae substructure was identified at a physical distance of less than 1 kpc.}
    \begin{tabular}{c|c|c|c|c|c|c|c}
    \hline
    GC ID & Substructure ID & $\mu_{\rm GC}$ [mag] & $\alpha$ [deg] & $\delta$ [deg]  & Potential Progenitor & Tidal Tail & Stream Name \\
    \hline
    \hline
NGC 6121 &     0                 &    11.71 & 245.90 & -25.47 &  Kraken          &        &             \\
NGC 3201 &         9             &    13.45 & 154.40 & -45.59 &  Sequoia/{\it Gaia}-Enceladus          &    G2/SF        &     \\
{\bf NGC 5139} &         3             &    13.58 & 201.70 & -46.52 &  {\it Gaia}-Enceladus/Sequoia          &  G1/SF          & Fimbulthul  \\
NGC 6171 &            16          &    14.03 & 248.13 & -12.95 &            &            &             \\
NGC 6266 &           19           &    14.16 & 255.30 & -29.89 &            &     G2       &             \\
{\bf NGC 5904} &         13             &    14.38 & 229.64 &   2.08 & Helmi streams/{\it Gaia}-Enceladus  & G1/SF        & \\
{\bf NGC 6362} &           15           &    14.40 & 262.98 & -66.95 &            &     G2       &             \\
E 3 &       31               &    14.54 & 140.24 & -76.72 &            &            &             \\
{\bf NGC 4590} &           17           &    15.06 & 189.87 & -25.26 &  Helmi streams          &     G1/SF       & Fj\"{o}rm             \\
NGC 6316 &         24             &    15.09 & 259.16 & -28.14 &            &            &             \\
{\bf NGC 1851} &         22             &    15.41 &  78.53 & -39.95 &  {\it Gaia}-Enceladus          &    G1/SF      &       \\
NGC 6934 &       27               &    15.97 & 308.55 &   7.40 &            &            &             \\
NGC 6981 &     25                 &    16.15 & 313.37 & -11.46 &   Helmi streams         &            &             \\
 IC 4499 &       29, 30               &    16.37 & 225.08 & -81.79 &  Sequoia          &            &             \\
 IC 1257 &       28               &    16.99 & 261.79 &  -6.91 &            &            &             \\
    \hline
    \end{tabular}
    \label{tab:GCs_of_interest}
\end{table*}

One of the benefits of employing hierarchical clustering is that we can select a context-dependent distance metric between objects. While it is possible that the intersection of the {\it Gaia} DR2 RR Lyrae variable catalogue and the APOGEE-2 survey's collection of absorption spectra\footnote{From the Sloan Digital Sky Survey website (\url{https://www.sdss.org/surveys/apogee-2/}): “The second generation of the Apache Point Observatory Galaxy Evolution Experiment
(APOGEE-2) observes the ‘archaeological’ record embedded in hundreds of thousands of stars to explore the assembly history and evolution
of the Milky Way Galaxy.” This data set contains information derived from spectroscopic measurements in the near-infrared.} is prohibitively small, a more sophisticated metric that incorporates stellar motions and chemical abundances is worth considering in future work. A potential improvement to this methodology would include a Milky Way phase space and metallicity vector ${\boldsymbol\chi = \left({\bf x}, {\bf v}, [{\rm Fe/H}]\right)}$, as well as a metric $D_{{\rm modified}}$ in a chemo-kinematic space dependent on the Galactic integrals of motion $E_{\rm tot}$ and ${\bf L}$:

\begin{align}
\label{eq:modified_metric} D_{{\rm modified}}(\boldsymbol\chi, \boldsymbol\chi')^{2} &\equiv  \beta_{0} (E_{\rm tot} - E_{\rm tot}')^{2} \\ \notag &+ \beta_{1}(L_{z} -  L_{z}')^{2} \\ \notag &+ \beta_{2}(L_{\perp} -  L_{\perp}')^{2} \\ \notag &+ \beta_{3}({\rm[Fe/H]} - {\rm[Fe/H]}')^{2},
\end{align}

\noindent where the total energy is $E_{\rm tot} = ({\bf v}\cdot{\bf v})/2 + \Phi({\bf x})$, the angular momentum is ${\bf L} = {\bf x}\times {\bf v}$, and $\{\beta_{i}\}$ are scaling factors that ensure each term has the appropriate units and weighted importance. Further investigation is needed to determine if this is a suitable metric or if different vector components should be introduced. Additionally, it is entirely possible that a large RR Lyrae dataset suitable for this modified metric on a Galactic scale is not yet available. In the future, infrared data collected by the
Nancy Grace Roman Space Telescope could be ideal for this type of analysis. The period-luminosity relation is, strictly speaking, only valid in the infrared (see \S \ref{sec:intro}), and this relation is what
makes RR Lyrae variable stars such reliable standard candles. Roman Telescope IR photometry and spectroscopic surveys could in principle be combined for more accurate RR Lyrae distance moduli (Equation \eqref{eq:absolute_magnitude}). Additionally, this would provide enough chemo-kinematic information such that the modified metric 
(Equation \eqref{eq:modified_metric}) might be tested and applied. The existence of a Galactic archaeology consortium and data analysis pipeline, as presented in \cite{ness2019}, would certainly help make a study of this kind possible.

An agglomerative, hierarchical clustering algorithm utilizing average linkage has time complexity ${\rm O}(n^{2} \, \log_{2} n)$ \citep{manning2008}, so applying particle-based clustering of this kind is probably ill-advised for samples bigger than
this one. Density-based hierarchical clustering, with time complexity ${\rm O}(n [\log_{2} n]^{3})$
, has been successfully employed
in astronomical contexts \citep{sharma2009, elahi2013, sanderson2015, sanderson2017}. Future work could be
devoted to applying such an algorithm to a larger catalogue such that the relevant instrument's selection function might be mitigated via random sampling.

Clustering algorithms are some of the most popular unsupervised learning processes; given the nature of astronomical
data sets (namely, RR Lyrae variables do not have name tags declaring to which globular cluster or dwarf galaxy progenitor it belongs) it is natural
to employ such processes. However, it may be possible to identify structures of interest using supervised learning once a sufficient number of training examples become available via observations or simulations.
For example, a random forest classifier \citep{breiman2001} could be employed, where labeled inputs (i.e., list of RR Lyraes known to belong to a particular globular cluster) pass through a forest of decision trees whose parameters are randomly chosen from the input’s data. The advantage of a random forest in this context is the incorporation of quantities not captured by an image, like metallicities or phase space coordinates. In the event such a random forest regimen is insufficiently accurate, implementing gradient boosting via XGBoost \citep{chen2016} would be a possible improvement.

\section{Conclusion} 
\label{sec:conclusion}

We have presented an analysis of the {\it Gaia} mission's RR Lyrae variable catalogue; the basis for this study was the repeated application of an agglomerative, hierarchical clustering algorithm to subsets of the variable star catalogue. The uncertainty in computed distance as a function of measurement errors in absolute/apparent magnitude and associated extinction is
considered, and we determined that the largest driver of distance modulus uncertainties results from large values of $\delta m_{V}$. Once
each RR Lyrae variable's 3D spherical coordinates were compiled (with the distance modulus used as a proxy for physical distance), we computed the condensed distance matrices for forty-eight catalogue subsets partitioned by HEALPix subregion. These matrices were used to create hierarchical clustering trees (with average linkage) containing archaeologically relevant substructures of RR Lyrae variables. The hierarchical clustering trees contain many
nearly redundant groupings at neighbouring clustering scales, so we proceeded in analyzing select scales
where with a local maximum of statistically significant groupings. The potentially interesting groupings were then compiled and analyzed.

Our results suggest that hierarchical clustering trees that use average linkage contain primarily stream extensions around globular clusters in a variety of Galactic environments. Substructures 0, 25, 29, and 30 are good candidates for making new connections to globular clusters important to the formation of the Milky Way; substructure 28 lies along the sightline near several known globular clusters, but has a distance modulus that suggests it is independent of these systems; substructure 31 is a possibly previously unknown satellite of the Large Magellanic Cloud. Follow-up studies would benefit from an exploration of the entire phase space \citep[e.g. Gaia EDR3 and DR3,][]{gaia2020}, as well as a sufficiently comprehensive mapping of the Milky Way that fills in any coverage gaps \citep[as the Rubin Observatory is expected to provide in the coming years,][]{najita2016}. Applying hierarchical clustering to three-dimensional data and validating with proper motions, however, is indeed effective in
identifying groupings of RR Lyrae variable stars, one of the undoubtedly reliable tracers of Galactic structure available
in the cosmos.

\section*{Acknowledgments}
\label{sec:acknowledgements}

This material is based upon work supported by the United States Air Force under Air Force Contract No. FA8702-15-D-0001. Any opinions, findings, conclusions or recommendations expressed in this material are those of the author(s) and do not necessarily reflect the views of the United States Air Force.

The authors wish to thank Anthony Brown, Marta Reina-Campos, and the anonymous referee for insightful comments that helped improve the manuscript.

BTC, KDS, and RM were supported by the Summer Research Program at MIT Lincoln Laboratory. BTC would like to thank the MIT Lincoln Laboratory staff, whose support during the internship made this work possible.

Part of this research was carried out at the Jet Propulsion Laboratory, California Institute of Technology, under a contract with the National Aeronautics and Space Administration.

This work has made use of data from the European Space Agency (ESA) mission
{\it Gaia} (\url{https://www.cosmos.esa.int/gaia}), processed by the {\it Gaia}
Data Processing and Analysis Consortium (DPAC,
\url{https://www.cosmos.esa.int/web/gaia/dpac/consortium}). Funding for the DPAC
has been provided by national institutions, in particular the institutions
participating in the {\it Gaia} Multilateral Agreement.

\vspace{2mm}

\underline{Software:} Numpy \citep{oliphant2007}, SciPy \citep{virtanen2020}, matplotlib \citep{hunter2007}, pandas \citep{mckinney2010}, astropy \citep{astropy2018}.

\vspace{2mm}

\underline{Data Availability Statement:} The data underlying this article are publicly available; they can be found using the citations and footnotes provided throughout the article. The article's online supplementary material includes a collection of this public data in a form that is well-suited for this study.


\bibliographystyle{mnras}
\bibliography{main_mnras} %




\appendix

\section{Mock RR Lyrae Streams from AMUSE Simulation}
\label{appendix:amuse}

In order to refine our RR Lyrae stream identification routine via an informed choice of $\tilde{X}_{\theta, {\rm min}}$, $R_{\rm max}$, (defined in \S\ref{subsec:hierarchical_clustering}), we must create a labelled dataset reflective of realistic Galactic dynamics and stellar populations. We select globular clusters NGC 362, NGC 2419, NGC 5466, Pal 5, and Pal 12 for further analysis using \cite{baumgardt2018, vasiliev2019}\footnote{\url{https://github.com/GalacticDynamics-Oxford/GaiaTools/}}. Each of these GCs have reported extratidal features and associations with with massive merger events in the Milky Way's formation history. This set has a diverse sampling of masses and galactocentric distances as well.

Globular clusters are notoriously difficult to model computationally, as the collisional nature of the internal dynamics necessitates using specialized gravity solvers. These $N$-body codes often carry time complexities on the order of ${\rm O}(n^{2})$; for $n\sim 10^{6}$, this can get prohibitively expensive. It is beyond the scope of this study to simulate the tidal stripping of globular clusters, so we propose the following simplified model of RR Lyrae stream creation. At each spatial location of the $i$th globular cluster, we create a particle of mass $M_{{\rm GC}, \, i}$ and calculate the relevant Jacobi radius \citep{binney2008}:

\begin{align}
\label{eq:jacobi} r_{J, \, i} &= |{\bf r}_{i}|\left(M_{{\rm GC}, \, i} \over 3 M_{\rm MW}(\bf{r}_{i})\right)^{1/3},
\end{align}

\noindent where ${\bf r}_{i}$ is the position vector of the $i$th globular cluster at the present epoch and $M_{\rm MW}(\bf{r}_{i})$ is the enclosed Galactic mass at this location. This is approximately the boundary at which the globular cluster and Galactic gravitational fields have equal influence. We then initialize 40 particles with an appropriate mass for RR Lyrae variables, $M = 0.65 \, M_{\odot}$ \citep{kolenberg2010}, and give them randomly distributed circular orbits about the globular cluster particle with a semi-major axis equal to the Jacobi radius at initialization time. If these RR Lyrae-like particles maintain orbits consistent with their natal GC, we should still be able to identify the remnant core using the $\max(N_{\rm RRL})$ value from \cite{sesar2013}. We initialize each GC in velocity space using representative Milky Way orbits from the {\tt galpy} package \citep{bovy2015}.

A natural runtime choice for this simulation is the globular cluster crossing time, $t_{{\rm cross} , i} \simeq |{\bf r}_{i}|/|{\bf v}_{i}|$. An order-of-magnitude estimate for the crossing time can be found using the enclosed Milky Way mass and assuming that the described model is virialized\footnote{Some of the GCs we are analyzing are extragalactic in nature, so this assumption should be used sparingly.}:

\begin{align}
t_{{\rm cross}, i} &= \sqrt{{|{\bf r}_{i}|^{3} \over G M_{\rm MW}({\bf r}_{i})}}.
\end{align}

In order to ensure that the stream gravity solvers adequately conserve energy (i.e., a fractional error ${\epsilon \lesssim 10^{-7}}$ throughout the simulation), we use the timestep ${\Delta t = 0.01 \, {\rm Myr}}$ and a simulation timescale ${t_{\rm end} = 35 \, {\rm Myr}}$ that is approximately the shortest crossing time (Pal 5). We simulate this system of massive particles with the aforementioned specifications using the AMUSE Python API \citep{portegieszwart2018, portegieszwart2013, pelupessy2013, portegieszwart2009b}; each stream is evolved forward in time using a symplectic (i.e., phase space conserving) integrator {\tt Huayno} \citep{pelupessy2012, janes2014} and bridged with a Milky Way-like (bar, disk, and bulge components) potential \citep{bovy2015}. Gravitational forces between streams are not considered, as they are negligible in comparison to background tidal forces.

\begin{figure*}
    \centering
    \subfloat[\centering The Jacobi radius (as defined in Equation \eqref{eq:jacobi}) for each of the five GC models as a function of simulation time. ]{{\includegraphics[width=8cm]{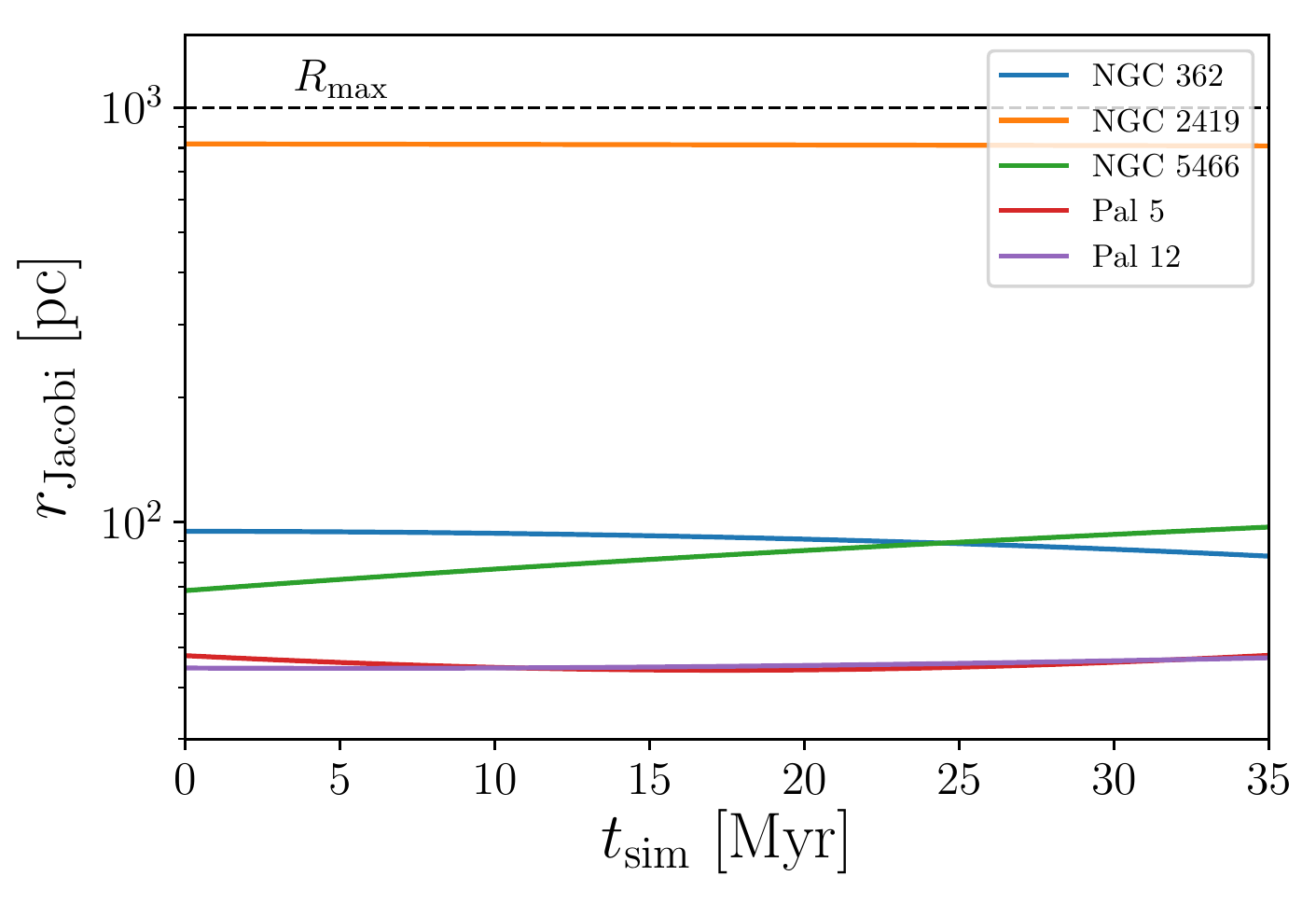} }}%
    \qquad
    \subfloat[\centering The median alignments (as defined in Equation \eqref{eq:dotProduct}) for each of the five GC models as a function of simulation time.]{{\includegraphics[width=8cm]{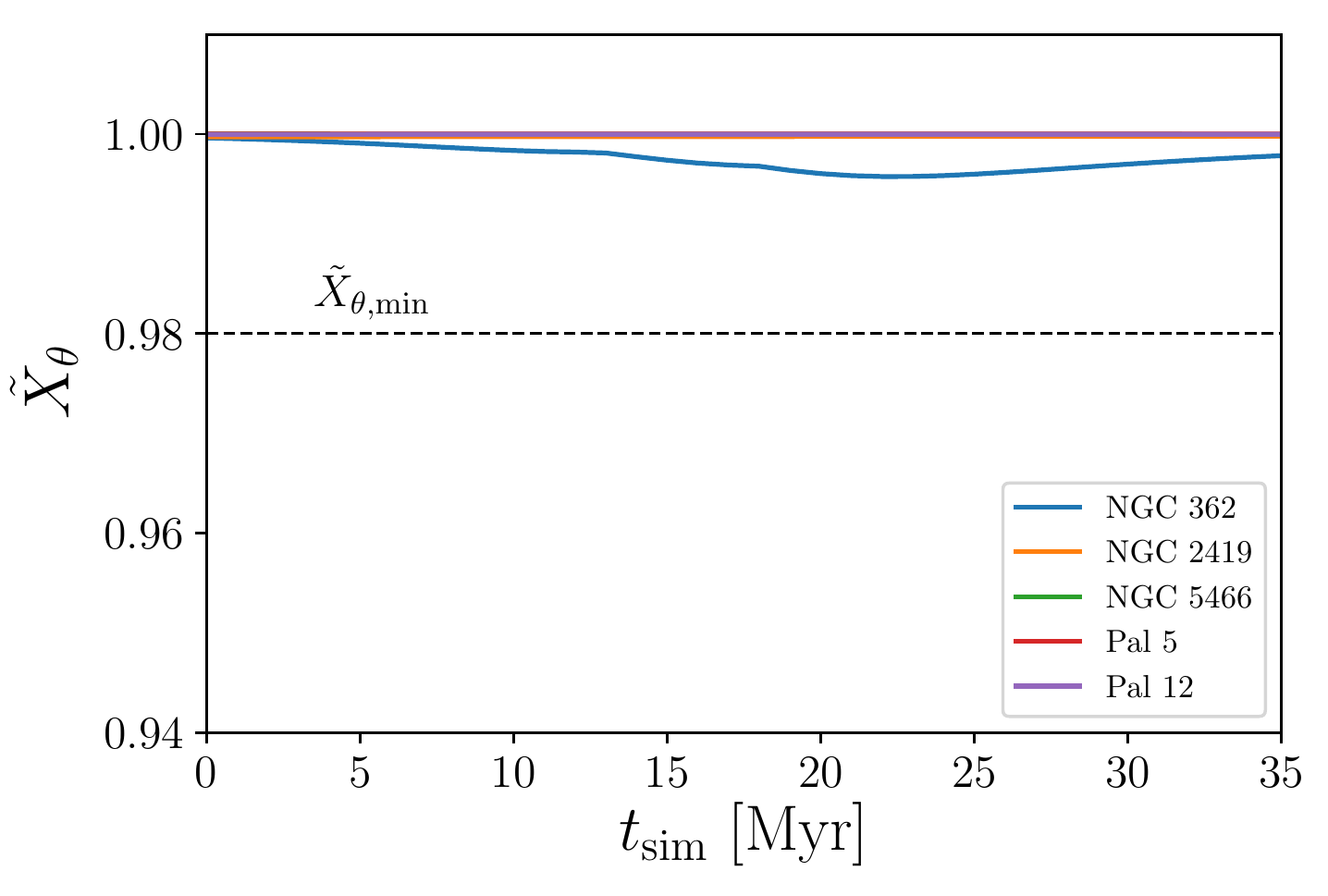} }}%
    \caption{Motivation for the choice of ($\tilde{X}_{\theta, {\rm min}}$, $R_{\rm max}$) stream parameters in the substructure identification routine presented in \S \ref{subsec:hierarchical_clustering}.}%
    \label{fig:streamParameters}%
\end{figure*}

Figure \ref{fig:streamParameters} shows the values of $\tilde{X}_{\theta}$ and the Jacobi radius as a function of simulation time for each of the five streams, as well as our stream parameter choice: {($\tilde{X}_{\theta, {\rm min}}$, $R_{\rm max}$) = (0.98, 1 kpc)}. 
We approximated the proper motion vectors (${\mu}_{\alpha}^{*}$, ${\mu}_{\delta}$) using ($v_{y}$, $v_{z}$) from the simulation data, where $y$, $z$ are part of the Galactic coordinate system defined in \S \ref{sec:methods}. The physical meaning of $\tilde{X}_{\theta, {\rm min}} = 0.98$ is that at least half of all stars in a retained grouping have proper motion vectors that deviate from the median grouping proper motion by $11.48^{\circ}$. This is a conservative threshold, as $\tilde{X}_{\theta}$ is very nearly equal to unity (i.e., proper motion deviations on the order of arcminutes rather than degrees) for all of the GC models during the simulation. Our choice of $R_{\rm max}$ ensures that two classes of substructure are retained: stellar streams extending an order-of-magnitude beyond the Jacobi radii of typical GCs, and atypical GCs like NGC 2419 that are more massive and further removed from the Galactic Center.

\section{Potentially extraneous substructure candidates}
\label{sec:potentially_extraneous_candidates}

Substructures identified in the hierarchical clustering forest described in \S \ref{sec:methods}, \S \ref{sec:results} were partitioned into two groups, where the attributes of the second group are provided in Table \ref{tab:rrl_substructures_appendix}. These substructures are related to those listed Table \ref{tab:rrl_substructures}, but for the sake of brevity we focused our analysis on the ones listed there and include the remainder in this appendix.

An illustrative example of this organization is the retention of substructure 3 for Table \ref{tab:rrl_substructures}, while listing substructures 2 and 4 are in Table \ref{tab:rrl_substructures_appendix}. All three substructures have sky plane coordinates consistent to within $0.1^{\circ}$ in both right ascension and declination with a similar number of RR Lyrae variables (15, 16, and 20, respectively). Each substructure is in the neighbourhood of the $\omega$ Centauri globular cluster, a known remnant of a dwarf galaxy merger whose tidal stream, Fimbulthul, was recently discovered using {\it Gaia} DR2 data. The Fimbulthul stream is separated from its progenitor by nearly twenty degrees on the sky plane and 1.5 kpc closer to Earth than $\omega$ Cen, thus making these candidate substructures unlikely members of the stream. Substructure 3 has the largest eccentricity, which means that if there is any connection to be made between $\omega$ Cen and the Fimbulthul stream via intermediate RR Lyrae variables \citep[$N$-body simulations suggest a stream of stars connecting the two should be present,][]{ibata2019}, substructure 3 is the most promising candidate.

\begin{table*}
    \centering
    \caption{Identified substructures where the constituent RR Lyrae variable sets are similar to the substructures listed in Table \ref{tab:rrl_substructures}. The hierarchical clustering process, and the substructure identification routine described in \S \ref{sec:methods}, do not strictly prohibit nearly redundant groupings.}
    \label{tab:rrl_substructures_appendix}
\begin{tabular}{c|c|c|c|c|c|c|c|c|c}
\hline
ID &  $\tilde{\mu}$ [mag] &          $\tilde{\alpha}$ [deg] &        $\tilde{\delta}$ [deg] &    $R$ [pc] &  $N_{\star}$ &  Eccentricity & Nearest GC & $d_{{\rm substruct} \to {\rm GC}}$ [mag] & $\Delta \mu (\tilde{\mu}, R=1 \, {\rm kpc})$ [mag] \\
\hline
\hline
1  &             12.77 &  245.9 & -26.5 &  138.48 &      19 &      0.994871 & NGC 6121 & 1.06 & 0.60 \\
2  &             13.81 &  201.6 & -47.4 &  274.07 &      15 &      0.979803 & NGC 5139 & 0.23 & 0.38 \\
4  &             13.84 &  201.6 & -47.4 &  264.15 &      20 &      0.887820 & NGC 5139 & 0.26 & 0.37 \\
5  &             13.99 &  154.4 & -46.4 &  388.76 &      29 &      0.977655 & NGC 3201 & 0.54 & 0.35 \\
6  &             14.00 &  154.3 & -46.4 &  239.44 &      23 &      0.985767 & NGC 3201 & 0.56 & 0.34 \\
7  &             14.01 &  154.4 & -46.4 &  160.30 &      17 &      0.978757 & NGC 3201 & 0.56 & 0.34 \\
8  &             14.11 &  154.4 & -46.4 &  126.51 &      19 &      0.946207  & NGC 3201 & 0.66 & 0.33 \\
10 &             14.38 &  229.6 &   2.1 &  779.15 &      26 &      0.998827 & NGC 5904 & 0.01 & 0.29 \\
11 &             14.38 &  229.6 &   2.1 &  779.15 &      26 &      0.998827 & NGC 5904 & 0.01 & 0.29 \\
12 &             14.39 &  229.6 &   2.1 &  276.70 &      23 &      0.986319 & NGC 5904 & 0.02 & 0.29 \\
14 &             14.58 &  263.0 & -67.0 &  490.72 &      19 &      0.999308 & NGC 6362 & 0.18 & 0.26 \\
18 &             15.01 &  189.9 & -26.7 &  240.83 &      16 &      0.992554 & NGC 4590 & 0.06 & 0.22 \\
20 &             15.13 &  255.3 & -30.1 &  442.13 &      30 &      0.923307 & NGC 6266 & 0.97 & 0.20 \\
21 &             15.13 &  255.3 & -30.1 &  258.84 &      27 &      0.948400 & NGC 6266 & 0.97 & 0.20 \\
23 &             15.44 &  255.3 & -30.1 &  262.99 &      19 &      0.997485 & NGC 6316 & 1.10 & 0.18 \\
26 &             16.30 &  313.4 & -12.5 &  278.98 &      32 &      0.932833 & NGC 6981 & 0.15 & 0.12 \\
\hline
\end{tabular}
\end{table*}

\bsp	
\label{lastpage}
\end{document}